\documentclass[aps,prb,twocolumn,showpacs]{revtex4}

\usepackage{graphicx}
\DeclareGraphicsExtensions{.png,.jpg,.eps}
\usepackage{multirow}
\usepackage{braket}
\usepackage{amsmath}
\usepackage[usenames,dvipsnames]{xcolor}

\begin{document}

\title{ Non-resonant electric quantum control of  individual on-surface spins}

\author{S. A. Rodr\'iguez$^{1}$, S. S.  G\'omez$^{1}$, J. Fern\'andez-Rossier$^{2}$
\footnote{On leave from Departamento de Fisica Aplicada, Universidad de Alicante, 03690 Spain },
A. Ferr\'on$^{1}$
}

\affiliation{ 
(1) Instituto de Modelado e Innovaci\'on Tecnol\'ogica (CONICET-UNNE) and 
Facultad de Ciencias Exactas, Naturales y Agrimensura, Universidad Nacional 
del Nordeste, Avenida Libertad 5400, W3404AAS Corrientes, Argentina.
\\(2) International Iberian Nanotechnology Laboratory (INL),
Av. Mestre Jos\'e Veiga, 4715-330 Braga, Portugal. 
}

\date{\today}

%%%%%%%%%%%%%%%%%%%%%%%%%%%%%%%%%%%%%%%%%%%%%%%%%%%%%%%%%%%%%%%%%%%%%%%%%%%%%%%%
\begin{abstract}

Quantum control techniques play an important role in manipulating and harnessing the properties of different quantum systems, including isolated atoms. Here, we propose to achieve quantum control over a single on-surface atomic spin using  Landau-Zener-Stückelberg-Majorana (LZSM) interferometry implemented with  Scanning Tunneling Microscopy (STM). Specifically, we model how the application of time-dependent, non-resonant AC electric fields across the STM tip-surface gap makes it possible to achieve precise quantum state manipulation in an isolated Fe$^{2+}$ ion on a  MgO/Ag(100) surface. We propose a protocol to combine Landau Zener tunneling with LZSM interferometry that permits one to measure the quantum spin tunneling of an individual Fe$^{2+}$ ion. The proposed experiments can be implemented with ESR-STM instrumentation, opening a new venue in the research of on-surface single spin control. 

\end{abstract}

%\pacs{73.22.Pr, 73.43.Cd}

\maketitle

The control of two-level quantum systems plays a central role in developing quantum technologies, such as quantum sensing and quantum computing.  Two complementary strategies can be used to achieve this goal. First, resonant excitation with either continuous wave perturbations or pulses\cite{Abragam_Bleaney_book_1970,vandersypen05,haroche06,hanson08}. Second, parametric manipulation of systems in the neighborhood of avoided crossings, that cause non-adiabatic transitions, which, together with quantum interference, lead to controllable state transitions\cite{nori1,nori2}. The protocol discussed is commonly known as Landau-Zener-St\"uckelberg-Majorana (LZSM) interference \cite{nori1,nori2}, and experiments in various systems have demonstrated its successful implementation. Notable examples include superconducting qubits\cite{oliver2009, SCQ1,SCQ2,SCQ3}, molecular nanomagnets \cite{LZqst,mm1}, semiconductor artificial atoms \cite{scaa}, impurity-based qubits \cite{ibq}, and NV centers \cite{nv}, among several others. In addition, LZSM interferometry has been extensively used for spectroscopy in different systems, including Josephson flux qubits \cite{oliver2009,ferron2010}. Is worth mentioning that Landau Zener (LZ) transitions have previously been employed to measure very small tunnel splittings in single-molecule magnets\cite{LZqst} and to demonstrate quantum state control in Ho, where the hyperfine interaction creates several avoided level crossings \cite{sugH2-2019quantum}.

 This work aims to lay the theoretical foundations of LZSM interferometry for individual surface spin controlled {\em electrically} with a Scanning Tunneling Microscope (STM). Our proposal requires instrumentation identical to state-of-the-art STM electron spin-resonance (STM-ESR) experiments\cite{hwang2022inst, Baumann2015}. There, the spin of the surface is driven, electrically\cite{Baumann2015,lado2017}, by a radio frequency voltage with frequency $f$. So far, it has been used to perform both 
 continuous-wave experiments and dynamical control with voltage pulses, with the frequency $f$ close to the resonance frequency $f_0$. The manipulation of the quantum state of an individual spin has been demonstrated\cite{willke2021coherent,sooyhon2023,wang23}. In our proposal, the driving frequency is far from resonance, and the driving mechanisms are thus different from previous works\cite{Baumann2015,willke2021coherent,wang23,willke2018,willke2018b,willke2019,willke2019b,yang2017,seifert2019,seifert2021,seifert2020} in ESR-STM.

\begin{figure*}[hbt]
\centering
\includegraphics[width=0.8\linewidth]{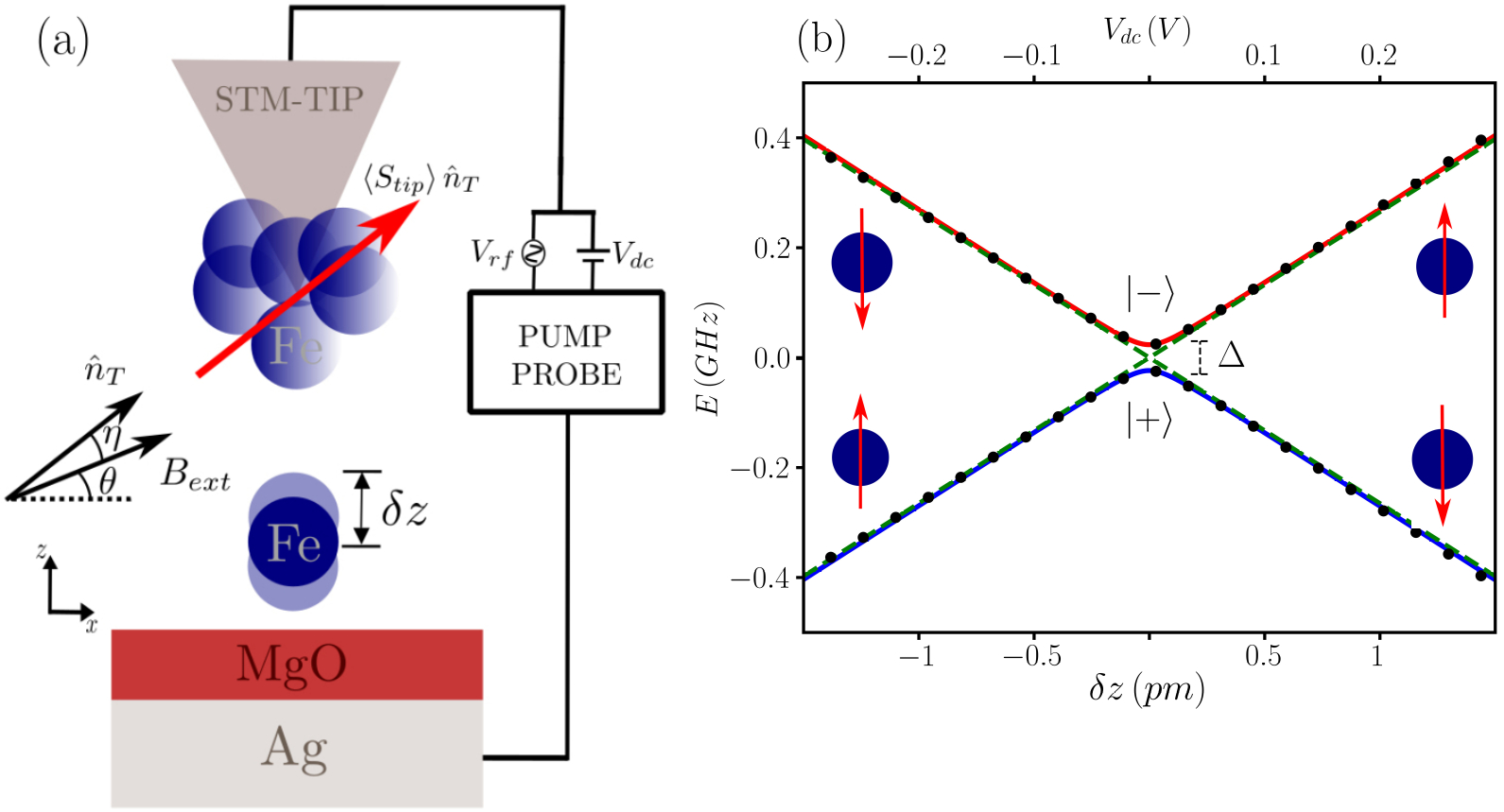}
\caption{\label{f1} 
(a) Sketch of the experimental setup: a spin-polarized STM drives the spin of a  single Fe$^{2+}$ ion on a MgO surface on top of silver. The external magnetic field forms an angle $\theta$ with the surface. Additionally, we can observe that the magnetization of the tip does not align with this field due to its anisotropy, forming an angle 
$\eta$ with the external field.
(b) Lowest energy levels of Fe$^{2+}$  as a function of the piezoelectric displacement or $V_{dc}$, as calculated with Hamiltonians Eq. (\ref{eq:TLS}) (dots) and Eq. (\ref{sm}) (continuous lines).}
\end{figure*}

 It is widely accepted\cite{Baumann2015,lado2017,Delgado2019} that the surface spin is coupled to the RF electric field associated with the RF voltage drop at the STM-surface junction. Piezoelectric displacement of the surface spin\cite{lado2017,yang2019,seifert2019} together with distance-dependent spin interactions between the surface spin and the magnetic tip provides a natural route for the coupling of the spin to the electric field. Here, we propose to leverage this piezoelectric coupling to carry out LZSM manipulation. We illustrate the application of our theory with the case of Fe$^{2+}$ on MgO, the physical system first studied with STM-ESR\cite{Baumann2015}. Figure \ref{f1}(a) shows the basic setup for the experiment, including the application of the magnetic field, the tip field generated by the Fe spin attached to the STM tip, and the Fe$^{2+}$ adsorbate on the MgO surface. Due to the crystal field of the surface and the atomic Spin-Orbit, Fe$^{2+}$ ground state is a non-Kramers doublet with S=2 and is expected \cite{baumann2015b,lado2017, Rodriguez2023} to feature quantum spin tunneling splitting $\Delta$, between two states with $S_z = \pm 2$, providing thereby a natural scenario for the existence of an avoided crossing. This feature is essential for employing LZSM interference to implement quantum control. We note that the analysis discussed in this work will apply to a wide class of non-Kramers doublets \cite{gatteschi06}, including atomic and molecular surface spins.

%check references
Many-body calculations\cite{Baumann2015,baumann2015b,lado2017,Rodriguez2023} show that the ground state doublet with $S_z=\pm2$ are separated  from the rest of the S=2 levels of Fe on MgO  by a $2.7$  THz  gap, to be compared with the   $\simeq 20$ GHz  ground state splitting. Therefore, we can consider Fe on MgO as an isolated two-level system with the states $| \pm 2 \rangle$. We then describe the ground state doublet of Fe$^{+2}$ on MgO in terms of a two-level system (TLS) (see Supplementary Material for calculation details):

\begin{equation}\label{eq:TLS}
H_{TLS}(t)=-24{\cal F}(t)\hat{\sigma}_x-h_{\rm eff}(t)  \hat{\sigma}_z 
\end{equation}
\noindent where $\hat{\sigma}_{x,z}$ are Pauli matrices acting on the subspace $S=2$, $S_z=\pm2 $, $\cal F$  is the coupling between the basis states, and $h_{eff}= 2 \mu_B g_zB_T$ is the effective zeeman field, due to the total magnetic field, with the anisotropic gyromagnetic constant $g_z$. $\cal F$ and $g_z$ are driven by the interplay of crystal field and spin-orbit coupling while the effective magnetic field results from the sum of three contributions \cite{Rodriguez2023,willke2019,seifert2020}: external magnetic field, stray field of the magnetic atoms in the tip, and exchange interaction with these atoms, treated semiclassically.

We now make the key observation that using the STM tip to control the system within the subspace of the ground state doublets is possible. We can perform such control with the application of a voltage across the tip-sample junction with a constant component, $V_{dc}$ and an oscillating component $V_{rf}(t)$, that lead to a piezoelectric distortion of the surface atom (see Fig. \ref{f1} (a)) \cite{lado2017,ferron2019,Rodriguez2023}:

\begin{equation}
%\delta z(t)\simeq\frac{q V_{dc}}{kd_{tip}}+\frac{q V_{rf}(t)}{kd_{tip}},
\delta z(t)\simeq\frac{q }{kd_{tip}}\left(V_{dc}+V_{rf}(t)\right),
\end{equation}

The effect of an electric field on the atom is the change in its position with respect to the surface. Changing the position of the atom modifies the influence of the surface on the atomic levels, and therefore both $\cal F$ and $h_{eff}$ will be affected. We have performed multi-configurational calculations of $\cal F$ and $g_z$ for different positions of the atom to obtain ${\cal F}(z)$ and $g_z(z)$ around the equilibrium position. Also, the magnetic field of the tip depends on the tip-atom distance, and therefore the piezoelectric displacement will affect $h_{eff}(z)$. To estimate the effect of the piezoelectric distortion, given the smallness of the displacement $\delta z$, we Taylor expand the Hamiltonian Eq. (\ref{eq:TLS}) keeping only the linear terms in $\delta z$, and therefore we write ${\cal F}= {\cal F}_{eq} + \alpha_{\cal F} \delta z$ and $h_{eff} = h^{eq}_{eff}  + \alpha_h \delta z$. According to our calculations,  $\alpha_h \simeq $270 GHz/nm and  $24\alpha_{\cal F} \simeq$ 1 GHz/nm. Since $\alpha_{\cal F}$ is much smaller than $\alpha_h$, it is safe to neglect the modulation of ${\cal F}$ from this point on. 
By changing the tip position in the experiment, we move the atom from its equilibrium position. Since the magnetic field of the tip is the sum of the dipolar field and the exchange term, we can find a tip-adatom distance, known as the NOTIN point \cite{seifert2020,Rodriguez2023}, where these two contributions cancel out. By also setting $B_{ext}=0$,  $h^{eq}_{eff}$ will be zero in this position. Oscillations will displace the atom from this position, producing a distorted field related to $\alpha_h \delta z$. As a result, we can write Hamiltonian Eq. (1) as:

\begin{equation}\label{sm}
H^s_{TLS}(t)=-\frac{\Delta}{2}\,\hat{\sigma}_x-\frac{\varepsilon(t)}{2}\, \hat{\sigma}_z 
\end{equation}

\noindent where the first energy scale, $\Delta = 48F_{eq}$, controlled by the quantum spin tunneling of the Fe$^{2+}$ ion, is known as \cite{nori1,nori2} tunneling amplitude. For zero-bias, we have  $\Delta \simeq 0.05$ GHz obtained with the value of ${\cal F}_{eq}$. The second term in Eq. (\ref{sm}) is known as the energy bias $\varepsilon(t) = 2\alpha_h \delta z(t)$. It is customary\cite{nori1,nori2}  to refer to the eigenstates of Hamiltonian Eq. (\ref{sm})  with $\Delta= 0$ as the diabatic states. In this system, they correspond to the eigenstates $S_z=\pm 2$. The adiabatic eigenvalues consist of the instantaneous eigenstates of the time-dependent Hamiltonian Eq. (\ref{sm}) for finite $\Delta$ and $\varepsilon$. These instantaneous eigenstates versus $\delta z$ or $ V_{dc}$ are depicted in Fig. \ref{f1}(b).
Black dots show the adiabatic eigenvalues for the Hamiltonian Eq. (\ref{eq:TLS}) and continuous lines were obtained for the simplified version Eq. (\ref{sm}). For small displacements (small voltages) calculations performed with the simplified version Eq. (\ref{sm}) agree with the exact solution obtained using Eq. (\ref{eq:TLS}). For larger values of voltage (bigger displacement of the atom) we observe some tiny differences between both solutions.

\begin{figure}[hbt]
\includegraphics[width=0.85\linewidth]{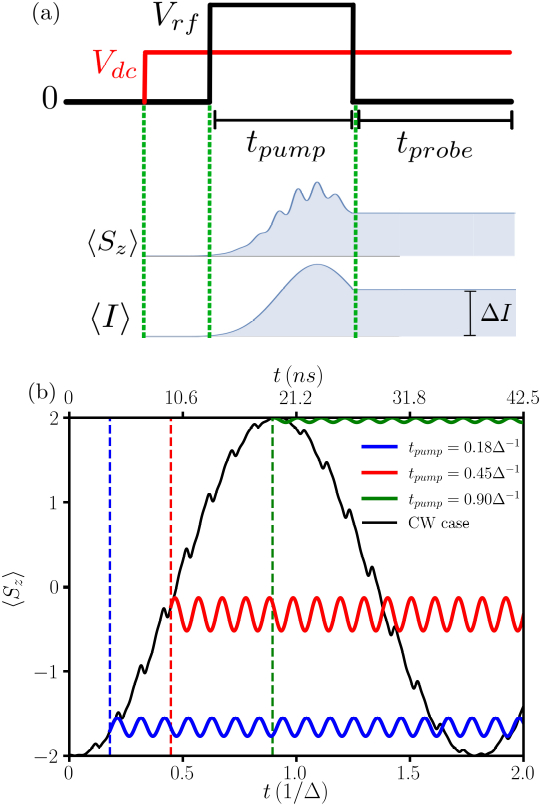}
\caption{\label{f2} (a) Sketch of the time dependence of voltage, Fe magnetization and current. Current flows in response to both DC and AC voltage. After  AC voltage train changes the Fe magnetization,
and, because of the tunnel magnetoresistance,  changes of DC current, making it possible to track the surface spin manipulation.
(b) Calculated time evolution of the average surface spin, $\langle S_z\rangle$, driven by the STM voltage (both DC and AC) in two different situations: cw-coherent driving (black lines) 
and pulsed-driving (colour lines), for which $AC$ driving is turned-off after an operation time $t_{pump}$. Fe magnetization oscillations when pump is off relate   to 
the free evolution in a superposition state. Calculations were performed for $f=10\Delta$, $V_{dc}=150$ mV, $V_{rf}=260$ mV and $B_{\rm ext}=0$.}
\end{figure}

We now illustrate how the spin orientation of the Fe$^{+2}$ ion on MgO can be controlled with pulses of AC voltages described with 
\begin{equation}
    V_{dc}+V_{rf}\sin{\omega t},
    \label{eq:RF}
\end{equation} where $ f= 2\pi \omega$ is the driving frequency, and $V_{rf}$ is the amplitude. 
The tip-atom distance is tuned as to be in the NOTIN point, where the tip magnetic field vanishes at the surface spin, leading to $\varepsilon=0$ in Eq. (\ref{sm})  when $V_{dc}=V_{rf}=0$ and $B^{z}_{ext}=0$. The three parameters that can be controlled experimentally are the amplitudes of the DC and RF voltages, $V_{dc}$, $V_{rf}$, and the frequency $f$. 
Numerical integration of the Schr\"odinger equation for Hamiltonian Eq. (\ref{eq:TLS}) yields the wave function that permits us to calculate the average off-plane magnetization of the Fe$^{2+}$ ion, $\langle S_z \rangle$. Our numerical simulations (see  Fig. \ref{f2}) show $\langle S_z\rangle$ as a function of time for fixed values of the frequency, $V_{dc}$ and $V_{rf}$ for four cases: continuous wave (CW) excitation, i.e., with $V_{rf}$ time-independent (black line), and AC pulses, where $V_{rf}$ is only finite during a finite interval of time with three different durations. In all cases, we have $f=10\Delta$, that is, completely out of resonance but within the limits of the current state of the art \cite{Baumann2015}. We find that, as long as the external voltage is on, the spin orientation is being controlled, in a time scale of 20 ns, determined by $\Delta$. Fig. \ref{f2}(b) illustrates the main point of this work: control over pulse duration permits to set the magnetization value at any arbitrary value\footnote{The small ringing oscillations of the magnetization once  $V_{rf}$ is switched off are due to are associated to the free evolution of a state linear combination of two different adiabatic states. The oscillation period  is controlled by $\sqrt{\Delta^2 +\varepsilon^2}$ (see color lines in Fig. \ref{f2} for $t>t_{pump}$)}. The suggested experimental protocol,  depicted in Fig. \ref{f2}(a),  should start by turning on $V_{dc}$ to initialize the system in a state where $S_z$ is close to either $+2$ or $-2$. The DC current measured before the RF pulse acts as a reference.  Application  the RF voltage pulse of duration  $t_{pump}$ changes the surface spin magnetization state, resulting in a different DC current once the AC pulse is over, on account of tunnel magnetoresistance \cite{Baumann2015}. 

\begin{figure}[hbt]
\includegraphics[width=0.95\linewidth]{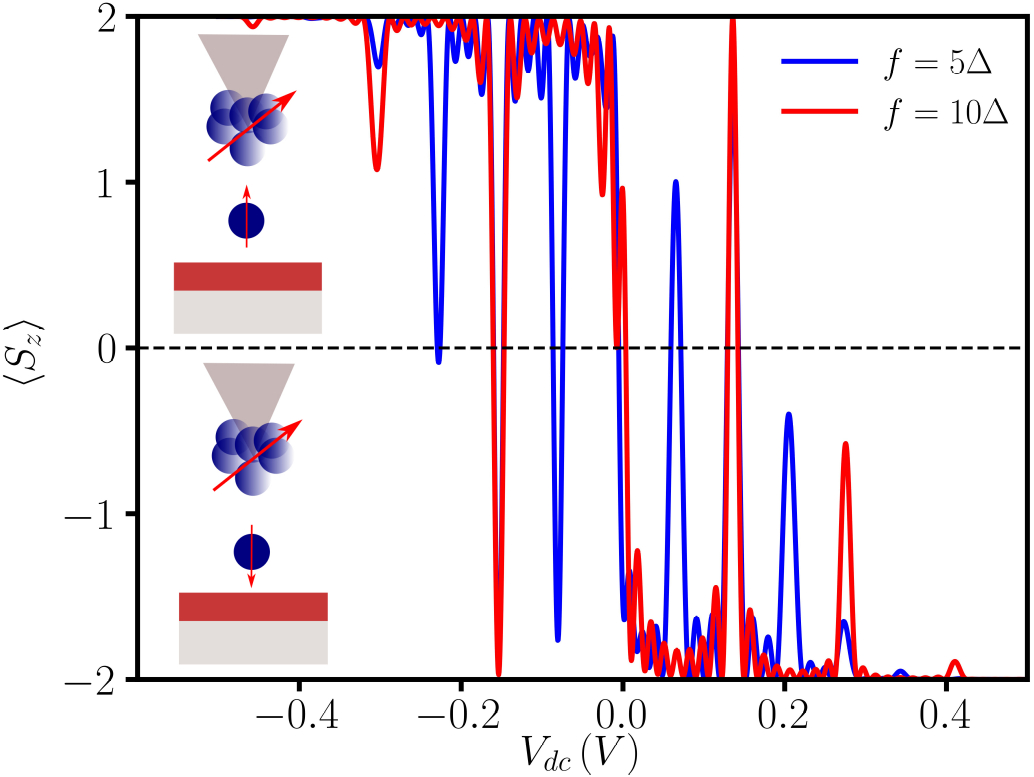}
\caption{\label{f3} $\langle S_z \rangle$ for different operation frequencies $f$ as a function of the detuning $V_{dc}$ for $V_{rf}=200$ mV, $t_{pump}=0.9\Delta^{-1}$ and $B_{ext}=0$.}
\end{figure}

To rationalize our results, we first discuss some limiting cases and review key concepts. We start analyzing a linear time-dependent perturbation $V(t)=v_s t$ where the velocity $v_s=\frac{dV}{dt}$ is the constant that controls the pace of the driving and induces, at the avoided crossing, the so-called Landau-Zener (LZ) transition. LZ transitions from the adiabatic ground state to the excited state occur with probability \cite{nori1} 

\begin{equation}
P_{LZ}=\exp{(-2\pi\delta_l)}=\exp{-\pi\left(\frac{\Delta}{\sqrt{2\hbar\frac{d\varepsilon}{dt}}}\right)^2}
\label{LZ}
\end{equation}

\noindent where $\frac{d\varepsilon}{dt}=\alpha_h q v_s/(z_n k)$.  This equation implies an exponential suppression of the transitions out of the adiabatic ground state as $v_s$ gets smaller, in agreement with the adiabatic theorem.

We now review the harmonic drive case Eq. (\ref{eq:RF}). Here, there is no general analytical solution, to the best of our knowledge. The dynamics can be approximated \cite{nori1,nori2} by adiabatic evolutions of the basis states far from the avoided crossing mediated by nonadiabatic  Landau Zener transitions at the avoided crossing (Adiabatic-Impulse Model\cite{nori1,nori2}). In this case, the $LZ$ transition probability, at a single passage, is given by the same Eq. (\ref{LZ}) with a modified expression for $\frac{d\varepsilon}{dt}=\alpha_h q v_{rf} /(z_n k)$ where $v_{rf}=V_{rf}\omega$ is the sweep velocity at the avoided crossing.
In the periodic driving case, explicit expressions for the occupation probability have been obtained in the fast or slow driving regime\cite{nori1,nori2,ferron2010}. Here we will consider the case of fast driving ($d\varepsilon/dt\gg \Delta^2$) and we are going to evaluate the transition probabilities. 

For $V_{dc}>0$ our system is started in the ground state, which has a a spin projection $S_z=-2$. We are interested in the probability of exciting transitions in 
the diabatic basis, i.e., the basis where Fe$^{2+}$ ion has $S_z=\pm 2$. Assuming that at $t=0$ the system is in the $S_z=-2$ state, 
the probability that the system switches to $S_z=+2$, in
the fast-driving regime is approximately given as \cite{nori1,nori2}
\begin{eqnarray}
P_{-2\rightarrow +2}(t)=\sum_n \frac{\Gamma_n^2}{2\Omega_n^2}\left(1-\cos{(\Omega_n t)}\right)   \nonumber \\
\Omega_n=\sqrt{\left(n\omega-\gamma V_{dc}\right)^2+\Gamma_n^2}\label{intlzs}
\end{eqnarray}
\noindent  where $\gamma=4\pi q\alpha_h/(k z_n)$, $\Gamma_n=\Delta J_n(\gamma V_{rf}/\omega)$ and $J_n$ is the Bessel function of the first kind.

One way to understand this equation is as follows: for every period of the RF frequency, the wave function undergoes a beam-splitter type of process, where it can either stay in the ground state or be excited.  As this beam-splitter process is repeated many times, given the oscillating nature of the driving potential, the final outcome requires the coherent sum of many passages, resulting in an interference pattern.

\begin{figure}[hbt]
\includegraphics[width=1.\linewidth]{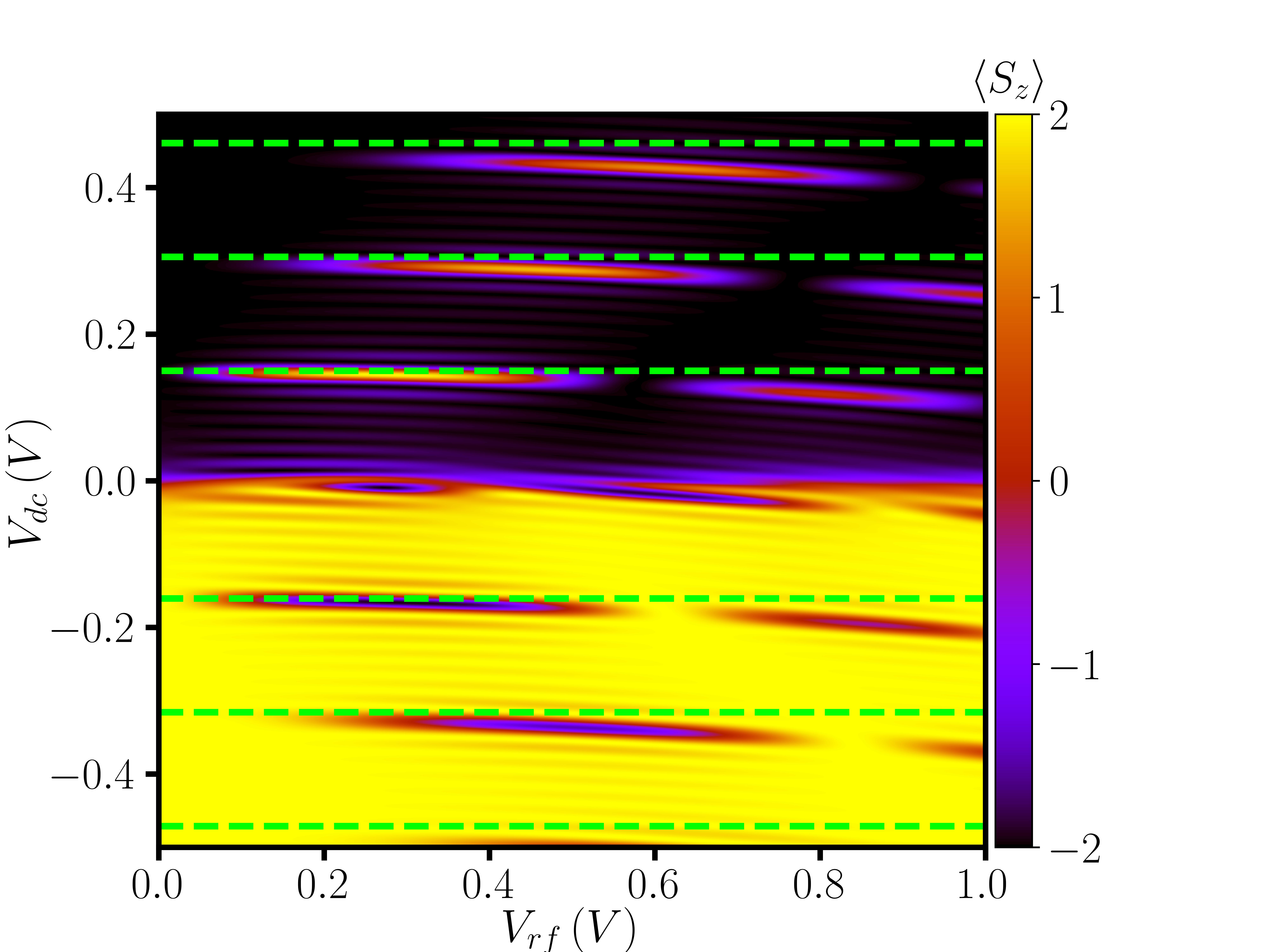}
\caption{\label{f4} Contour map of $\langle S_z\rangle$ as a function of the detuning $V_{dc}$ and the driving amplitude $V_{rf}$ for $t_{pump}=0.9\Delta^{-1}$ and $f=10\Delta$.
}
\end{figure}

We now discuss our systematic exploration of how different parameters that can be changed in an experiment can be used to optimize the control over the spin state and to explore its physical properties. In Fig. \ref{f3} we show the $V_{dc}$ dependence of the magnetization, measured right at the end of a pulse of fixed duration ($t_{pump}= 0.9 \Delta^{-1}$), for two different frequencies.
From  Eq. (\ref{intlzs}) it is apparent that the probability of transitions in the diabatic basis is a sum of periodic functions whose amplitude is maximal whenever $n\omega=\gamma V_{dc}$, giving a resonance pattern, as obtained in our numerical simulations using Eq. (\ref{eq:TLS}). Notably, there are two values of $V_{dc}$ ($V_{dc}\simeq \pm 0.15V$) for which the spin flipping is complete. Since our model neglects spin relaxation and decoherence, the validity of our predictions will hold as long as the operation time is significantly smaller than the decoherence time  $T_2$. For Fe$^{2+}$ on MgO, 
experiments show\cite{Baumann2015}   $T_2\simeq 210 \pm 50$ ns, larger than control times reported here.

We now address the influence of the amplitude of the driving AC voltage, $V_{rf}$ on the LZSM spin control process.  From Eq. (\ref{intlzs}) we see that $V_{rf}$ enters $\Gamma_n$ that controls the amplitude of the different harmonics. Thus, for certain values of $V_{rf}$ the Bessel functions, and therefore the amplitudes, vanish. This is seen in the phase diagram
of Fig. \ref{f4}, where we plot a contour map for the spin projection of the adatom as a function of the $V_{dc}$ and $V_{rf}$, measured at  $t=t_{pump}=0.9\Delta^{-1}$, with a fixed
driving frequency $f=10\Delta$. We also plot the position of the resonances obtained using Eq. (\ref{intlzs}) as horizontal green dashed lines. 
Here, we can appreciate interference patterns that exhibit fringes that rise from the constructive interference between successive Landau–Zener transitions. 
The agreement between our numerics and Eq. (\ref{intlzs}) is very good for the first resonances. The horizontal modulation of the amplitude of the resonances reflects the  $V_{rf}$  dependence of $\Gamma_n$ anticipated in eq  (\ref{intlzs}).
We also note that, in contrast with eq. (\ref{intlzs}) our numerical calculation is not symmetric with respect to the inversion of the sign of $V_{dc}$. This difference arises from the fact that Eq. (\ref{intlzs}) is derived assuming a static ${\cal F}$ and linearized $h_{eff}$, whereas in our calculation, we do include their complete dependence on $\delta z$, and therefore on $V_{dc}$.

To conclude,  we propose a novel approach to obtain the Quantum Spin Tunneling (QST) of the Fe$^{2+}$ ion, combining LZ and LZSM. We note that LZ was used to determine QST in single-molecule magnets, where the prefactors that relate  $\delta_l$ in eq. (\ref{LZ})  are known.  In our case   $\delta_l$  depends both on $\Delta$ and $\alpha q v_s/ z_n k$. Importantly, the resonance pattern of the LZSM permits one to infer
 $\alpha q v_s/ z_n k$. Combining both LZ and LZSM it should be possible to determine the quantum spin tunneling splitting $\Delta$. It is important to note that LZ protocol is not efficient to carry out control over the spin magnetization because it requires a very small sweeping rate and therefore operation times much longer than those of LZSM (see SM for details).
 
In summary,  we have proposed an approach to control the spin of individual magnetic adatoms using ESR-STM instrumentation that, unlike previous work\cite{},  operates non-resonant pulses, leveraging both on LZ 
and LZSM  mechanisms associated to avoided crossings in the spectrum.   Specifically, we have studied in detail the paradigmatic case of Fe$^{+2}$ ion on MgO\cite{Baumann2015}. The proposed LZSM approach offers several significant advantages over the conventional resonant control method.
%
%\begin{enumerate}
 First, the LZSM mechanism is much faster. As evident from Fig. \ref{f2} and Eq. (\ref{intlzs}), the time scale for controlling the orientation of the surface spin is determined by the inverse of tunneling splitting $\Delta$. This contrasts with conventional resonant driving, where the time scale is controlled by the Rabi coupling $\Omega$. For instance, Supplemental Material of Reference [\onlinecite{yang2019coherent}] shows that, for $V_{RF}=400$ mV, the Rabi oscillations for Fe$^{2+}$ on MgO have their first maximum at $t \simeq 60$ ns, three times slower than LZSM protocol (see  Fig. \ref{f2}). We note that there is room for much faster maniupulation as $\Delta$ can be dramatically larger in other non-Kramers doublets. However, magnetization control could only occur if $\varepsilon$ can be made larger than $\Delta$, which deserves further study. Second, the LZSM approach  is still effective for much smaller values of $V_{RF}$,  as shown in Figs. \ref{f3} and \ref{f4}, where the flipping process still  occurs for $V_{RF}\simeq 200$ mV. At this $V_{RF}$ value, experiments using conventional ESR-STM \cite{yang2019coherent} show no Rabi flip using the resonant mechanism, presumably because $\Omega T_2$ is not large enough. Finally, unlike the resonant mechanism, the LZSM mechanism does not require fine-tuning the frequency (see Fig. \ref{f3}). 

We now discuss briefly the limitations of the LZSM. First, our simulations neglect the effect of decoherence, so that their validity is only warranted provided that the pulses are much shorter than $T_2$, which sets a lower limit for the frequency and an upper limit for $t_{\rm pump}$. Our simulations assume a pure state at $t=0$. Therefore, temperature should be significantly smaller than the energy splitting in Fig. 1 (b), for finite $\delta z$. For a splitting in the range of $1$ GHz, this requires  temperatures in the range of  $T \leq 40$ mK. ESR-STM has been demonstrated at this low temperatures\cite{Akaje2021mk}, although in most cases higher temperatures are used. Finally, we assumed that the nuclear spin of Fe is zero, which is correct for  the most abundant isotope ($^{56}$Fe), but not for $^{57}$Fe ( $2.1\%$ abundance) that has a nuclear spin $I=1/2$ that would add a $A I_z \sigma_z $ term\cite{willke2018b} to the Hamiltonian and change the results of our simulations \cite{willke2018b}.

In conclusion, the LZSM protocol presents a promising alternative to conventional resonant control, offering faster operation, lower driving frequencies, effectiveness at lower driving amplitudes, and reduced sensitivity to frequency tuning. In addition, by combining LZ and LZSM we have shown that it is possible to determine the magnitude of the quantum spin tunneling splitting $\Delta$.

\vspace{0.5cm}

We thank Philip Willke, Andreas Heinrich and María José Sánchez for useful discussions.
A.F. acknowledges ANPCyT (PICT2019-0654). A.F., S.S.G. and S.A.R.
acknowledge CONICET (PUE22920170100089CO and PIP11220200100170) and 
partial financial support from UNNE.
J.F.R.  acknowledges financial support from FCT (Grant No. PTDC/FIS-MAC/2045/2021),
SNF Sinergia (Grant Pimag,  CRSII5\_205987), Generalitat Valenciana funding Prometeo2021/017
and MFA/2022/045, and  funding from MICIIN-Spain (Grant No. PID2019-109539GB-C41).

\bibliography{biblio}{}

\end{document}